# MULTI LAYERS SUPPLY CHAIN MODELING BASED ON MULTI AGENTS APPROACH

Samia Chehbi, Yacine Ouzrout, Aziz Bouras
*{samia.chehbi, youzrout, abouras}@univ-lyon2.fr*

*This paper proposes a strategic multi layers model based on multi agents approach for supply chain system. It introduces a formulation and a solution methodology for the problem of supply chain design and modeling. In this paper we describe and analyze the relationships among main entities of a supply chain, such as suppliers, producers, and distribution centers, in the aim to design the agents and define their behavior. We also study, how these relationships can be formulated in a multi layer model. Finally, a generic multi agent model is illustrated.*

## 1. INTRODUCTION

The most popular research topic in the field of supply chain (SC) management is the formulation of strategic and efficient model. This can be opted by different manners, by using artificial intelligence tools or integer programming methods (Wu, 2001). The problem is commonly arises in the evaluation of some parameters characterizing SC state. A number of production producers supply a collection of distribution centers with multiple products, which, in turn, supply customers with specified demand quantities of different products. The challenge is to determine the number, location, capacity, and type of convenient actors to minimize the total cost of the SC. The mathematical problem of formulation in production context exists since a long time where some works, like the one of Goeffrion and Graves (Goeffron, 1972) described a multi-commodity single-period production-distribution problem and solved it by *Benders Decomposition*. Recently, Hong Y. et al. (Hong, 2003) has developed a proved method based on constraints to design a strategic production-distribution model. Other works have been published recently under this theme, like (Dong, 2003), where Dong J. et al. analyze the formulation and design a demonstrated mathematical model based on lemmas and theorems. Most efforts in these works consider SC activities separately and proceed by studying SC as a linear model and try to represent it globally. We provide another view to model SC, considering it as a non linear system with a high level of complexity, and we try to apply technical tools, classically used to resolve complex systems design and modeling. In the next sections, we describe briefly some SC features, then, we detail our formulation steps and describe the parameters, variables and constraints used to design the multi layers system. After, we proceed by giving some key issues in SC design and modeling using multi agent systems and finally, we give an example of dimensioning supply chain problem.

## 2. FORMULATION OF SUPPLY CHAIN ORGANISATION

Today, competition in global markets with heightened expectations of customers has pushed the enterprises to invest and make more importance of their SCs. Consequently, to reduce cost and improve service levels, effective SC strategies must take into account interactions at the various levels in the SC. To improve their SC performance, firms must focus on understanding most information and relationships nature of all the partners and SC actors. Having an idea of a model projecting their SC, these firms can improve many strategic decisions like forecasting operations, predicting customers' demands and decreasing warehouse stocks costs.

This paper shows how theoretical programming formulations can be applied to SC design problem, and focuses on the fact of dividing SC into upstream and downstream parts by considering the manufacturer as the reference mark. The main objective in the proposed approach is to evaluate an objective function (to maximize) and a cost function (to minimize). We assume that the SC's flow concerns a single family of product.

Actors in classical SC models (Simchi-Levi, 2000) are not organized, and interactions are ignored. This leaded us to propose in a precedent work (Chehbi, 2003) a multi-layers reorganization in order to facilitate the SC evaluation's phase. Hence, any SC can be transformed into a multi-layers architecture (figure 1).

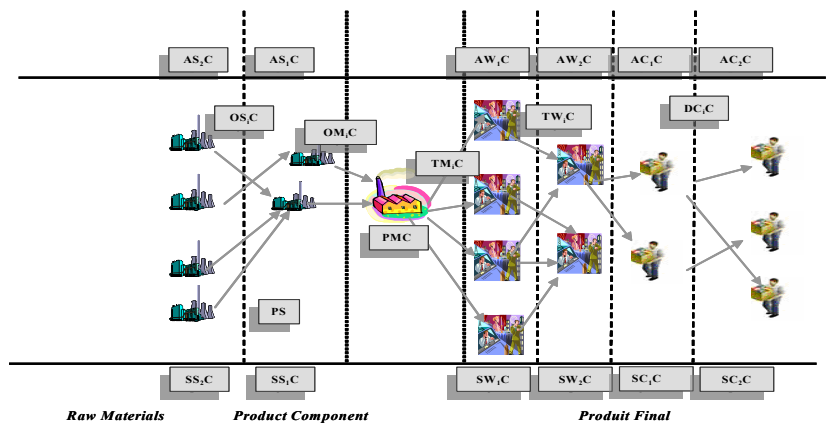

Figure 1 – Multi Layers Supply Chain Architecture

### 2.1. Supply Chain Parameters

Before formulating the model, we introduce the basic parameters notations and definitions. In this study, we use the following indices: $s \in S$, a set of candidate suppliers; $M$ the single manufacturer; $w \in W$, a set of warehouses and distribution centers; $c \in C$, a set of customers; $pc \in PC$, a set of product components needed for production; $p \in P$, the single type of product characterizing the SC. Using these notations, we describe some considered costs as follows:

*Added Cost*: It is the cost obtained when introducing an actor in the chain. Hong et al. (Hong, 2003) used the same principle in their formulation of a logical constraints model for SC. They call it the fixed cost to open and operate an actor. Each actor has its proper *added cost* in the SC. Relating to each actor type, we distinguish: *ASiC* (*Added Supplier 'i' Cost*), *AWiC* (Added Warehouse *'i'* Cost) and *ACiC* (Added Customer *'i'* Cost).

*Action Costs*: Signifies the internal cost evaluated for each actor. We distinguish two types of costs; the first is the *Production Cost* of one unit of the final product in the case of the main manufacturer (*PMC*), or a unit of a product component in the case of a supplier $PS_i$. The second type concerns the *Storage Cost* of a unit of the final product or its components (*SS$_i$S for suppliers, SW$_i$C for warehouses and SS$_i$C for customers*).

*Interaction Costs*: Interactions between supply chain actors play an important role in the total cost. Along the upstream supply chain, we define *TMiC* (Transportation Cost Between the manufacturer and its '$i^{\prime th}$' customer), *TWi*C (Transportation Cost between warehouse*s*) and *DCiC* (*Distribution Cost* between customers). In the downstream chain, we define *OMiC* as the cost of materials ordered by the main manufacturer to its $i^{th}$ supplier and *OSiC* between manufacturers to deliver the product.

## 2.2. Evaluation of objective function

Before providing the total cost function, we define other notations related to each actor location in the multi layer architecture. A supplier $i$ located in a level $n$ is indicated by $S_{Ln}^n$ with $L_n$ the number of suppliers in the layer $n$. The same rule is applied to the other actors, so we have as a result the indices matrix (*Supply Chain Matrix*) described below.

$$\begin{bmatrix} S_1^n & \ldots & S_1^1 & W_1^1 & \ldots & W_1^p & C_1^1 & \ldots & C_1^m \\ S_2^n & \ldots & S_2^1 & W_2^1 & \ldots & W_2^p & C_2^1 & \ldots & C_2^m \\ S_3^n & \ldots & S_3^1 & W_3^1 & \ldots & W_3^p & C_3^1 & \ldots & C_3^m \\ \ldots & \ldots & \ldots & \ldots & \ldots & \ldots & \ldots & \ldots & \ldots \\ S_{L_n}^n & \ldots & S_{L_1}^1 & W_{L_1}^1 & \ldots & W_{L_p}^p & C_{L_1}^1 & \ldots & C_{L_m}^m \end{bmatrix}$$

Using these notations and all parameters evolved before, we give the total cost of supply chain as the sum of *added*, *action* and *interaction costs*. By evidence, the *total added cost* is the sum of *added supplier costs*, *added warehouses costs* and *added supplier costs*.

$Total\ Added\ Supplier\ Cost = \sum_{i=1}^{L1} ASiC + \sum_{i=1}^{L2} ASiC + \ldots + \sum_{i=1}^{Ln} ASiC,\ Total\ Added\ Warehouse\ Cost = \sum_{i=1}^{L1} AWiC + \sum_{i=1}^{L2} AWiC + \ldots + \sum_{i=1}^{Lp} AWiC,\ and\ Total\ Added\ Customer\ Cost = \sum_{i=1}^{L1} ACiC + \sum_{i=1}^{L2} ACiC + \ldots + \sum_{i=1}^{Lm} ACiC$

After evaluating *added costs* of each type of supply chain actors, we sum them to have the total added cost in the chain. To evaluate *total action cost*, it is important to notice that there are probably actors which are not producers or storage centers; we interpret this by introducing a binary coefficient $q_{Lj}^i$ to the supplier *i* belonging to the $j^{th}$ layer, whether it is a producer { $q_{Lj}^i = 1$ } or not { $q_{Lj}^i = 0$ }.

*Total Production Cost = Production Costs of suppliers + Production Cost in the main manufacturer and Total Storage Cost = Storage Cost of suppliers + Storage Cost of warehouses + Storage Cost of customers.*

$$\text{Total Action Cost} = \sum_{i=1}^{L1} PS_i \cdot q_{L1}^i + \sum_{i=1}^{L2} PS_i q_{L2}^i + \ldots + \sum_{i=1}^{Ln} PS_i q_{Ln}^i + PMC + \sum_{i=1}^{L1} SS_i C \cdot q_{L1}^i + \sum_{i=1}^{L2} SS_i Cq_{L2}^i + \ldots + \sum_{i=1}^{Ln} SS_i Cq_{Ln}^i + \sum_{i=1}^{L1} SW_i C \cdot q_{L1}^i + \sum_{i=1}^{L2} SW_i Cq_{L2}^i + \ldots + \sum_{i=1}^{Lp} SW_i Cq_{Lp}^i + \sum_{i=1}^{L1} SC_i C \cdot q_{L1}^i + \sum_{i=1}^{L2} SC_i Cq_{L2}^i + \ldots + \sum_{i=1}^{Lm} SC_i Cq_{Lm}^i$$

*Total Action Cost = Sum of ordered costs between suppliers + sum of ordered costs between the main manufacturer and its suppliers + sum of transportation costs between the main manufacturer and its customers + sum of transportation costs between warehouses + sum of distribution costs between customers.*

$$\text{Total Interactin Cost} = \sum_{i=1}^{L2} OS_i C + \ldots + \sum_{i=1}^{Ln} OS_i C + \sum_{i=1}^{L1} OM_i C + \sum_{i=1}^{L1} TM_i C + \sum_{i=1}^{L2} TW_i C + \ldots + \sum_{i=1}^{Lp} TW_i C + \sum_{i=1}^{L1} DC_i C + \sum_{i=1}^{L2} DC_i C + \ldots + \sum_{i=1}^{L(m-1)} DC_i C$$

By consequence, the objective function is given by: *Min F = Total Added Supplier Cost + Total Added Warehouse Cost + Total Added Customer Cost + Total Action Cost + Total Interaction Cost.*

## 3. MULTI AGENT SYSTEMS MODELING

This section presents an issue for modeling the dynamic behavior of the proposed SC multi layers model. Our aim is that to obtain an efficient tool of simulation which can be applied to quantify the flow of SC information. With this described model, we think be capable to determine strategic policies are effective in smoothing and reducing variations in the SC. In most recent works in this topic, SC and enterprises networks have been a fertile area of multi agent simulations. That's because there is a growing need to developing decentralized efficient tools aiding to more performed management tools.

Referring to (Ferber, 1995), a multi agent system is a collection of, possibly heterogeneous, computational entities, having their own goals and problem-solving capabilities. Won et al. (Dong, 2002) suggest a set of interactive agents for Harbor SC network. Lin et al. (Lin, 1998) present multi agents architecture to model and simulate SC information system, they propose a shared environment based on agents simulating orders processes. Researches on agents-based SC management can be divided into three types: (1) Agent-based architecture for coordination, (2) agent-based simulation of SCs and (3) dynamic formation of SCs by agents. Our current work is a combination between the two first types of researches. It proposes an agent-based architecture doted of decision making agents to insure collaboration between SC parts. Based on various designs for multi agent systems in the literature (Dong, 2002) and many previous researches, we try to design an agent-based SC model, described statically and dynamically via three types of agents (figure 2). An agent type called '*controller*' to model the SC dimensioning, a set of agents to model SC dynamic, divided into physical agents representing tangible existing objects (such as wholesalers, customers, etc) and logical agents defining a virtual agent for each layer. They are doted of information functions used to control the information flows and manage the interactions as described bellow:

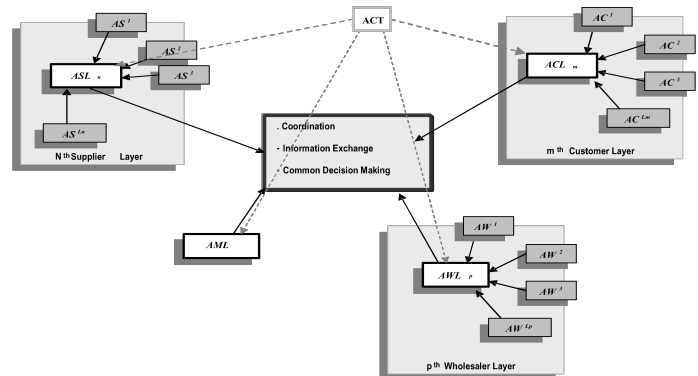

Figure 2 – Multi-Layers Agent SC Model

### 3.1 Agent-Actors

For each SC entity, we define a specified agent. So, $AC_j^i$ means the $i^{th}$ customer agent in the $j^{th}$ customer layer. Applying the same signification, we have $AS_j^i$ for suppliers and $AW_j^i$ for wholesalers. They are designed with classical standardized functionalities to communicate, negotiate and send or receive requests from other agent-actors of the same layer and the agent-layer (see figure 3, (a)). Three principle modules are implemented in the agent-actor; each one insures a specific task: The communication module serves as a reception-sending filter with the other agents of the same layer. There may be several types of information exchanged, such as products demand, negotiation messages, asking for a shared information, etc. The knowledge management module contains all parameters perceived by the agent and a part of the database. For example, action costs are the most important variables existing in the internal database. Moreover, this agent has the possibility of asking for external data from other agents. To coordinate between previous modules, a coordination module is added.

Each layer is managed by an agent, responsible for the management of the agents of its layer. We define one agent for the manufacturer (*AMM*), one for each customers layer (*ACM$_i$*), one for each suppliers layer (*ASM$_i$*) and also one for each wholesalers *layer (AWM$_i$)*. These agents have to insure and maintain the minimization cost of their layers when they receive the order from the controller agent (Figure 3, (b)). Interactions are always made between agents of to adjacent layers. In addition to the existing modules implemented in the Agent-Actors, we define the "Decision Making" module which manages the negotiations between agents.

### 3.3 Agent-Controller

We define a unique controller-agent in the system to evaluate strategic decisions for supply chain dimensioning. All the formulation part is implemented in the decision making module of the controller. The controller communicates continuously with all the agents-layers of all the chain in order to update its information. The design of the controller is illustrated below (figure 3, (c)).

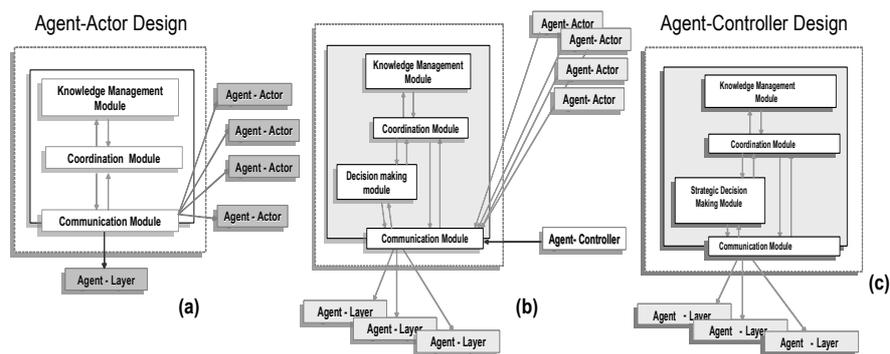

Figure 3 – Agents design

## 4. ILLUSTRATIVE EXAMPLE

In this section, we present an example of a simplified supply chain illustrated in (figure 4). The chain is related to the product *P*, assembled and delivered by the main manufacturer *Man*. In the manufacturing process, we distinguish an intermediate product component *IP* assembled and delivered by a secondary center of assembly *D*. There are three first suppliers *{A, B, C}* for three raw materials *{M1, M2, M3}*, one storage center *E* for the product component and the raw material *M3* in addition to three storage centers *{G, L, M}* for the final product *P*. We also have six final customers *{N, O, P, Q, R, S}*.

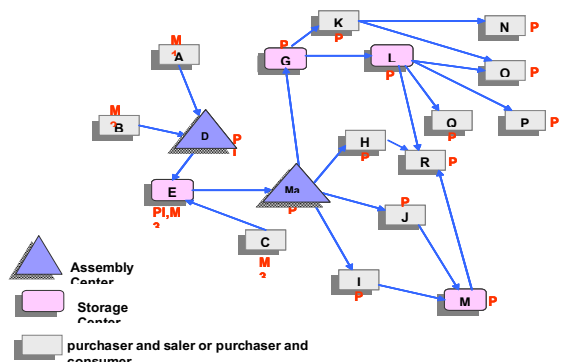

Figure 4 – SC description – Illustrative Example

### 4.1. Multi Agent Model

In order to decompose the chain into layers, it is important to extract a hierarchic tree to distinguish and define the various levels. We take as reference mark the manufacturer of the final product and then we advance in the hierarchy in the two directions (customers and suppliers) (figure 5). In this example we can divide the chain into six layers *{S1, S2, S3, C1, C2, C3}*.

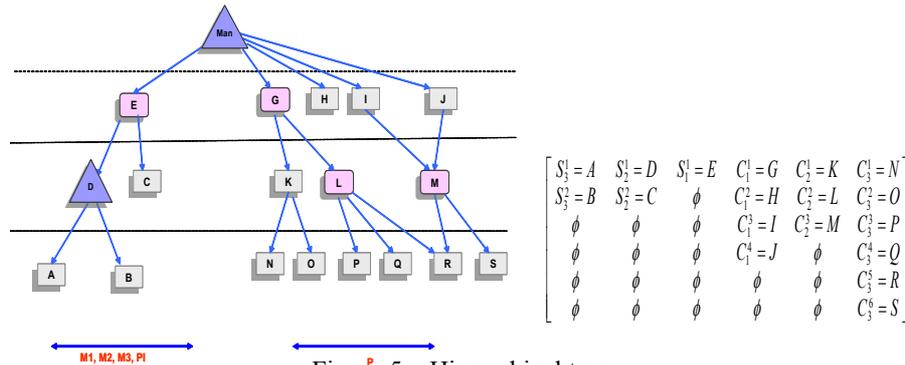

$$\begin{bmatrix} S_3^1 = A & S_2^1 = D & S_1^1 = E & C_1^1 = G & C_2^1 = K & C_3^1 = N \\ S_3^2 = B & S_2^2 = C & \phi & C_1^2 = H & C_2^2 = L & C_3^2 = O \\ \phi & \phi & \phi & C_1^3 = I & C_2^3 = M & C_3^3 = P \\ \phi & \phi & \phi & C_1^4 = J & \phi & C_3^4 = Q \\ \phi & \phi & \phi & \phi & \phi & C_3^5 = R \\ \phi & \phi & \phi & \phi & \phi & C_3^6 = S \end{bmatrix}$$

Figure 5 – Hierarchical tree

Each element in the matrix is represented by one reactive agent containing its actor's information, and each column in the matrix is represented by one cognitive agent manager to coordinate between layers and make partial common decisions, in addition to another cognitive agent defined to take strategic decisions for supply chain dimensioning. By consequence, our multi agent organization contains seven reactive agents related to six cognitive agent-managers in each layer and one cognitive agent-controller. In order to clarify the use of each agent, we describe a simple example of a scenario showing the problem of supply chain dimensioning.

### 4.2. Dimensioning problem description

We assume that the main manufacturer *Man* is located in France; its storage center *G* is located in USA and it has to know if it is profitable to open a secondary center of assembly in USA instead of delivering products to USA with a high cost of transportation. We suppose that in the state $St_1$, the agent controller *AC* has estimated the total cost of the chain at a value $CSt_1$. We suppose also that the state of the chain in the case of adding the secondary center of assembly in USA is called $St_2$, were the manufacturer buys product components *{PI, M3}* in USA and assembles them in the added center (figure 6).

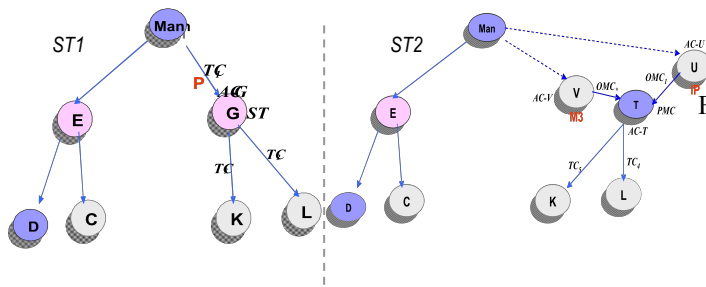

Figure 6 – Illustration of dimensioned supply chain problem

Each agent-actor sends to the agent-manager of the specific layer the values of estimated added costs *AC* of the actors *{U, V, T}* respectively, in addition to the production cost of the secondary manufacturer *T {PMC}*. The agents-managers, in turn, send these information in addition to the interaction costs *{OMC$_1$, OMC$_2$, TC$_4$, TC$_5$}* to the agent controller. This one collects all cost values, calculates the total cost *CSt$_2$*, compares with the previous cost *CSt$_1$* and finally decides which state is profitable for the manufacturer; the state *St$_1$ (CSt1<CSt2)* or *St$_2$ (CSt$_1$>CSt$_2$)*.

## 5. CONCLUSION

Although there is a wealth of literature and research on modeling of strategic supply chain, there is an apparent lack of theoretical consideration of supply chain constraints. In this paper, we introduced a strategic supply chain formulation model based on ordered layers and including different constraints. The complexity of a supply chain and its details leaded us to use a "complex systems" based design, aiming at the minimization of a global system cost while satisfying customers demand. We have explained the considered parameters, variables and constraints and presented a multi-layers model (an organization of non linear sub-systems). We also proposed a multi-agent implementation for the used formulation. This implementation is still in progress and a simple example is given at the end of the paper to help understanding the used organisation.